\documentclass[twocolumn,showpacs,amsfonts,amssymb,amsmath,floats,aps]{revtex4-1}
\usepackage{graphicx}
\usepackage{dcolumn}
\usepackage{multirow}

\usepackage{braket}

\def\w{\omega}

\def\tr#1{\mbox{Tr}\left[#1\right]}

\usepackage[normalem]{ulem}
\usepackage{upgreek}
\usepackage{bbold}

\def\doubleunderline#1{\underline{\underline{#1}}}
\usepackage{xcolor}

\begin{document}

\title{Long-time coherence in fourth-order spin 
correlation functions}

\author{Nina Fr\"ohling}
\address{Lehrstuhl f\"ur Theoretische Physik II, Technische Universit\"at Dortmund,
Otto-Hahn-Stra{\ss}e 4, 44227 Dortmund, Germany}

\author{Frithjof B.\ Anders}
\address{Lehrstuhl f\"ur Theoretische Physik II, Technische Universit\"at Dortmund,
Otto-Hahn-Stra{\ss}e 4, 44227 Dortmund, Germany}

\date{\today}

\begin{abstract}

We study the long-time decay of
fourth-order electron spin correlation functions for an isolated singly charged
semi-conductor quantum dot. The electron spin dynamics is governed by
the applied external magnetic field as well as the hyperfine interaction.
While the long-time coherent oscillations in the  correlation functions 
can be understood within an semi-classical approach treating the Overhauser field as frozen,
the field dependent decay of its amplitude reported in different experiments
cannot be explained by the central-spin model indicating the insufficiency of such a description.
By incorporating 
the nuclear Zeeman splitting and the strain induced nuclear-electric quadrupolar interaction, 
we find the correct crossover from a fast decay in small magnetic fields
to a slow exponential asymptotic in large magnetic fields. It originates from a competition between
the quadrupolar interaction inducing an enhanced spin decay and the nuclear Zeeman term
that suppressed the spin-flip processes. 
We are able to explain 
the magnetic field dependency of the characteristic long-time decay time $T_2$
depending on the experimental setups. The calculated asymptotic values
of $T_2 = 3 -4\,\upmu$s  agree qualitatively well with the experimental data. 
\end{abstract}

\maketitle

\section{Introduction}

The spin of a  single electron  or a hole confined in a semiconductor
quantum dot (QD) is a promising candidate for
the realization of solid state based quantum bits
\cite{HansonSpinQdotsRMP2007,Glazov2013:rev,SchliemannKhaetskiiLoos2003,Greilich2006,GreilichBayer2007}.
In contrast to  defects in diamonds \cite{JelezkoWachtrup2004,NVreview2006}, such 
QDs can be easily integrated into conventional semiconductor
devices and allow ultrafast optical preparation and
control \cite{Greilich2006,GreilichBayer2007}.
A main challenge in such devices is the loss of information due to spin
decoherence. The high localization of the electron wave function
in the QD reduces all decoherence facilitated by free electron motion, but simultaneously increases the hyperfine interaction strength
between the confined electron spin and the surrounding
nuclear spins. This hyperfine interaction dominates
the short-time dynamics of the confined electron spin
\cite{Merkulov2002,CoishLoss2004,FischerLoss2008,Testelin2009}.

The analysis of the spin-noise spectrum
\cite{Crooker2009,Crooker2010,Dahbashi2012,Glazov2012,PhysRevB.89.045317,Sinitsyn2012,
PhysRevLett.115.207401,Glasenapp2016}, the Fourier
transform of the second-order spin autocorrelation function $C_2(t)$,
reveals some of the intrinsic dynamics of the central
spin interacting  with its environment. The short-time dynamics of $C_2(t)$ can be understood 
using a semi-classical approximation \cite{Merkulov2002,ChenBalents2007,Glazov2012} 
where the Overhauser field generated by 
the nuclear spins is treated statically due to the separation of time scales.
The short-time
dephasing of the electronic spin is caused by averaging  the electronic spin precession 
over a Gaussian distribution of Overhauser fields defining a characteristic time scale $T^*\approx 1-4$\,ns.
While the remaining part of $C_2(t)$ is partially
protected against further decay by conservation laws \cite{UhrigHackmann2014} within the central-spin model \cite{Gaudin1976} in the absence of an external magnetic field, it has been suggested \cite{Sinitsyn2012,PhysRevLett.115.207401,Glasenapp2016} that the  strain-induced nuclear electric quadrupolar interaction
plays a crucial role in understanding the long-time
spin dynamics in QDs. Randomly orientated nuclear quadrupolar easy axis \cite{Bulutay2012}
leads to a breaking of  the conservation of total spin inducing a second decoherence time $T_H$
of the order to $200-800$\,ns in typical QDs \cite{PhysRevLett.115.207401} independent
of sign of the charge in the QDs. 
Cywi\ifmmode~\acute{n}\else \'{n}\fi{}ski
et al.\ \cite{PhysRevLett.102.057601} investigated the electron spin 
decoherence function by
applying a canonical transformation to the Gaudin model \cite{Gaudin1976} valid in large external
magnetic fields.
More in depth reviews on the spin dynamics in a single QD can be found
in Refs.\ \cite{HansonSpinQdotsRMP2007,Glazov2013:rev}.

Recently, higher-order spin autocorrelation functions have been promoted \cite{LiuSham2010,Bechtold2016,LiSinitsyn2016}  
as an indicator for quantum effects not accessible by  $C_2(t)$.
Bechtold et al.\   have measured the joint probability $w(t_1,t_2)$ 
of still finding a spin-down state
in two consecutive measurements at the waiting times $t_1$ and $t_1+t_2$ after 
preparing the spin of a singly charged  quantum dot in the spin-down state. 
Using quantum measurement theory \cite{QuantumMeasurement1995},
we are able to link   $w(t_1,t_2)$ to a fourth-order spin correlation function.
Press et al\ \cite{Press2010} have addressed the question of prolonging
electron spin coherence in a QD via a three pulse spin echo method.
It turns out that the measured probability function is likewise given by
a fourth-order spin correlation function. A long time decoherence in a similar order of magnitude was also observed in mode-locking experiments \cite{Varwig2016}.
While  the spin decay 
is related to a weak coupling Markovian process parametrized with a phenomenological decay rate
in the literature \cite{LiuSham2010,Bechtold2016,LiSinitsyn2016}, this
paper aims for an understanding of the experimentally observed response
based on a microscopic Hamiltonian 
and the accurate evaluation of the quantum mechanical traces of such a 
coupled electron-nuclear system.

The differences and similarities between these two experiments 
\cite{Bechtold2016,Press2010} will be expanded upon in this paper.
In particular, we will show that the absolute values as well
as the magnetic field dependency of the long-time
decay time $T_2$ of the order of $1-4\,\upmu$s depend on the type of experiments.
Spin-echo protocols  \cite{Hahn1950} have been used for a long 
time to extend the spin coherence time \cite{Viola1998,Greilich2009} in QDs.
We analyze in detail the physical origin of the different time scales observed in the experiment. In particular, we will argue and provide strong numerical evidence that the
additional magnetic-field dependent 
long-time scale $T_2$ \cite{Bechtold2016}  is related to
the interplay between  nuclear-electric quadrupolar interaction causing the long-time decoherence
and the nuclear Zeeman effect which suppresses the decoherence mechanism:
Without the quadrupolar interaction, quantum coherence would be maintained as predicted by the semi-classical approximation (SCA)\cite{Merkulov2002,ChenBalents2007,Glazov2012} in contradiction to the experiments  \cite{Press2010,Bechtold2016}.
Our theory
provides an explanation for threshold behavior of $T_2$ 
as function of the external magnetic field observed in experiments.

\subsection{Plan of the paper}

We study the fourth-order correlation function using three different
methods: 
(i) the SCA of a Gaussian
distributed frozen Overhauser field \cite{Merkulov2002,ChenBalents2007,Glazov2012},
(ii) the exact enumeration of quantum-mechanical expectation values
in finite size system based on the exact diagonalization
of the Hamiltonian, and 
(iii) an iterative Lanczos based approach \cite{Lanczos50aniterative}
to the real-time dynamics of $w(t,t)$ that
is numerically very expensive but required to estimate finite size corrections
to $T_2$. In order to ascertain what interactions influence the long-time behavior of the
fourth-order correlation function, the full quantum-mechanical model is required, incorporating
both nuclear-electric quadrupolar couplings and Zeeman splitting of
the nuclei.

%

The rest of the paper is structured as follows. 
In section \ref{sec:quantum-measurement}, we derive the relation
between the joint probability $w(t_1,t_2)$  and a fourth-order correlations function using 
quantum measurement theory \cite{QuantumMeasurement1995}.
In addition we calculate the probability of finding the electron spin again in an spin ground state
after applying a pulse sequence of $\pi/2-\pi-\pi/2$ pulse on that  spin ground state.
We will introduce the central-spin model (CSM) in Sec.\ \ref{sec:CSM}
and
the additional nuclear-electric quadrupolar  interaction in Sec.\ \ref{sec:Hq}.
Section \ref{sec:methods} is devoted to the applied methods and explicitly discusses
the evaluation of the fourth-order correlation function using exact diagonalization (ED)
in Sec.\ \ref{sec:ED} while we provide details of the Lanczos approach to $w(t,t)$
in Sec.\ \ref{sec:lanczos-restart}.

The results will be presented in Sec.\ \ref{sec:results}.
We start with a comparison between
$w(t_1,t_2)$ obtained using ED and the SCA in Sec.\ \ref{sec:CMS-SCA-ED} with the CSM.
Sec.\ \ref{sec:CMS-HQ}  covers to the influence of the nuclear-electric quadrupolar interaction
as well as the nuclear Zeeman term onto the fourth-order correlation function. 
By combining $H_{\rm CSM}$ 
with quadrupolar interaction and Zeeman splitting of the nuclei,
we present our final results for $w(t_1,t_2)$  in Sec.\ \ref{Finley_full}
and show the very good agreement with the experiments.
In Sec.\ \ref{sec:spin-echo} we extent our theory to the spin-echo experiments \cite{Press2010}.
At the end, we close with a summary and a conclusion.


\section{Quantum measurement and fourth-order correlation}
\label{sec:quantum-measurement}

\subsection{Three measurement pulse experiment}
 \label{4corr_3pulse}
 
The experiment conducted by Bechtold et al.\ \cite{Bechtold2016} measures the probability of the central spin being in a spin $\ket{\downarrow}$ state both at time $t_1$ and time $t_1+t_2$,
after a  pump pulse transferred the electron spin of 
the singly negatively charged InGaAs quantum dot  into a $\ket{\downarrow}$-state. 
In this section we show that this probability can be casted into a fourth-order autocorrelation function for the spin projector 
$P_\downarrow = \ket{\downarrow}\bra{\downarrow}$, 
which is related to a sum of second and fourth-order spin correlation functions in the high temperature limit.

We describe these three consecutive projection processes
 through quantum measurement theory \cite{QuantumMeasurement1995}.
The first probe at time $t_1$ transfers the density operator from $\rho_{\text{init}}$ to
\begin{align}
\rho(t_1)=\frac{1}{w_1(t_1)}P_{\downarrow} U(t_1) \rho_{\text{init}} U^\dagger(t_1) P^{\dagger}_{\downarrow},
\end{align}
where the time evolution operators is given by $U(t) = \exp(-\text{i}H t)$ and
\begin{align}
w(t_1)=\braket{P_{\downarrow}(t_1)}_{\text{init}}
\end{align}
is the probability  to measure the electron spin $\ket{\downarrow}$ state at time $t_1$. 
We use the notation $\braket{\hat O}_{\text{init}}=\text{Tr}(\hat O \rho_{\text{init}})$ where $\hat O$ can be any operator,
as well as $P_{\downarrow}= P_{\downarrow}^\dagger = P_{\downarrow}^2$.
 The conditional probability  of measuring the electron spin $\ket{\downarrow}$ state again at time $t_1+t_2$ is given by
\begin{align}
w(t_1| t_2)=\text{Tr}\left(P_{\downarrow}U(t_2)
\rho(t_1) U^\dagger(t_2) P^\dagger_{\downarrow} \right).
\end{align}
Therefore, the joint probability function for both measurements can be written as
\begin{align}
\begin{split}
w(t_1, t_2)&= w(t_1 | t_2)w(t_1)\\
&= \text{Tr}\left(P_{\downarrow}U(t_2)
P_{\downarrow} U(t_1) \rho_{\text{init}} U^\dagger(t_1) P^{\dagger}_{\downarrow}
 U^\dagger(t_2) P^\dagger_{\downarrow} \right)
 \\
&=\braket{P^{\dagger}_{\downarrow}(t_1)P^{\dagger}_{\downarrow}(t_1+t_2)P_{\downarrow}(t_1)}_{\text{init}}.
\end{split}
\end{align}
Due to the nature of the first pump pulse, the initial density operator is given by  $\rho_{\text{init}}=2P_{\downarrow}(0)/D$ in the high temperature limit, with $D$ being the dimension of the Hilbert space, 
and we arrive at
\begin{align}
\label{eq:w-t1-t2-def}
w(t_1, t_2)& =\frac{2}{D}\text{Tr}\left(P_{\downarrow}(t_1)P_{\downarrow}(t_1+t_2)P_{\downarrow}(t_1)P_{\downarrow}(0)
\right).
\end{align}

Since the experiments are usually performed at temperature around $T=4-6\,$K and the hyperfine interaction strength
corresponds to about 50\,mK, employing the  high temperature limit is well justified.

It turns out that the
joint probability function is a special case of a general fourth-order 
autocorrelation function defined
as 
\begin{eqnarray}
G_4(t_1, t_2, t_3)
&=& \frac{2}{D}\text{Tr}\left(P_{\downarrow}(t_1)P_{\downarrow}(t_2)P_{\downarrow}(t_3)P_{\downarrow}(0)\right)
\label{eqn:g4}
\end{eqnarray}
such  that  $ w(t_1, t_2)= G_4(t_1, t_1+t_2, t_1)$

Using the identity $P_{\downarrow}= \mathbb{1}/2-S_z$ and the high-temperature limit, we obtain
\begin{eqnarray}
G_4(t_1, t_1+t_2, t_1) &=& \frac{1}{4}+C_2(t_1)+C_2(t_2)+\frac{1}{2}C_2(t_1+t_2)
\nonumber \\
& &+2C_4(t_1, t_1+t_2, t_1)
\label{eq:G4-C2-C4}
\end{eqnarray}
with  second-order spin autocorrelation functions 
\begin{align}
C_2(t) = \braket{S_z(t)S_z}
\end{align}
and the general unsymmetrized fourth-order autocorrelation function
\begin{align}
C_4(t_1, t_2, t_3) = \braket{S_z(t_1)S_z(t_2)S_z(t_3)S_z}.
\end{align}
All expectation values are  calculated with respect to the high-temperature density operator 
$\rho_{\text{ht}}=\mathbb{1}/D$.

The properties of the second-order autocorrelation function $C_2(t)$ are well understood
for singly charged semiconductor quantum dots
 \cite{Glasenapp2016, GreilichBayer2007, PhysRevLett.115.207401, Sinitsyn2012} and are experimentally accessible via spin noise measurements as well as direct measurement of the real-time dynamics \cite{Bechtold2016}.
In a strong external magnetic field applied in the $x$-direction, 
$C_2(t)$ decays to zero for times $t \gg T^{\ast}$ \cite{Testelin2009, PhysRevB.89.045317},
where $T^{\ast}$ denotes the time scale defined by the fluctuation of the Overhauser field. 
This allows us to discuss possible long time limits of $G_4$. The maximum of the fourth-order spin autocorrelation function $C_4$ is $1/16$, since correlation is highest if the spin is in the same state at all times. Therefore, long-time limits for $G_4$ will lie between $1/4$ (full decoherence of $C_4$) and $3/8$ (maximum coherence of $C_4$).

\subsection{Spin-Echo measurements}
\label{spin-echo}

In a recent ultrafast optical spin echo experiment \cite{Press2010}, in a single quantum dot, an intrinsic long-time decoherence scale
$T_2$ has been determined by initializing the electron spin in the QD in the ground state in an external magnetic field in the $x$-direction and then applying a $\pi/2-\pi-\pi/2$ pulse sequence with a fixed duration
of $2T$ between the two $\pi/2$ pulses and the $\pi$-pulse at a time $T+\tau$. The $\pi/2$-pulse
rotates the electron spin into the $z$-direction where it starts precessing with the Larmor-frequency $\w_L$.
After the time $T+\tau$, the spin component perpendicular to the external magnetic field is flipped.
In a system of pure static dephasing via a frozen distribution of local magnetic fields, there would be
a revival of the signal at $2T$ for $\tau=0$. By varying $\tau$, interference oscillations can be observed
where the amplitude is taken as a measure for restoring quantum coherence via spin-echo pulses.
For a large magnetic field of $2-10$\,T, values of $T_2\approx 2.6\,\upmu$s have been reported \cite{Press2010}.

The probability amplitude for finding the electronic spin again in the spin ground state
$\ket{g_0}$
after the application of  the pulse sequence for two fixed nuclear spin 
configurations $\vec{m},\vec{m'}$ is given by
\begin{eqnarray}
A(T,\tau,\vec{m},\vec{m'}) &=& \bra{g_0,\vec{m}} 
U_y(\pi/2) e^{-\text{i}H(T-\tau)} 
\\
&&\times U_y(\pi) e^{-\text{i}H(T+\tau)} U_y(\pi/2)
\ket{g_0,\vec{m'}}
\nonumber
.
\end{eqnarray}
Since the nuclear spin configurations are completely undetermined by the experiment,
we need to sum over all nuclear contributions in the probability function and arrive
at
\begin{eqnarray}
P_{g_0g_0}(T,\tau) &=& 
\frac{2}{D}\sum_{\vec{m},\vec{m'}}  
|A(T,\tau,\vec{m},\vec{m'}) |^2
\nonumber
\\
&=&
\frac{2}{D}\sum_{\vec{m}}
\bra{g_0,\vec{m}}
U^\dagger_y(\pi/2)
e^{\text{i}H(T+\tau)}
U^\dagger_y(\pi) 
\nonumber\\
&&\times
e^{\text{i}H(T-\tau)} 
U^\dagger _y(\pi/2)
\hat P_{g_0}
U_y(\pi/2) e^{-\text{i}H(T-\tau)} 
\nonumber\\
&&\times U_y(\pi) e^{-\text{i}H(T+\tau)} U_y(\pi/2)
\ket{g_0\vec{m}}.
\end{eqnarray}
Here $D/2$ denotes the number of nuclear spin configurations
and the sum runs over all possible nuclear configurations.
Defining the projector onto the electron spin ground state,
\begin{eqnarray}
\hat P_{g_0} &=& \ket{g_0}\bra{g_0},
\end{eqnarray}
and
\begin{eqnarray}
A&=& U_y(\pi/2)\hat P_{g_0} U^\dagger _y(\pi/2)
\\
\bar A &=& U_y^{\dagger}(\pi/2)\hat P_{g_0} U_y(\pi/2)
\\
B &=& U_y(\pi) 
\end{eqnarray}
yields another fourth-order correlation function
\begin{eqnarray}
P_{g_0,g_0}(T,\tau) 
&=& \frac{2}{D} \tr{  B^\dagger  \bar{A}(T-\tau) B A(-T-\tau) }.
\end{eqnarray}
Identifying $t_1=T+\tau$ and $t_2=T-\tau$, reveals the similarity to $w(t_1,t_2)$ introduced
in Eq.\ \eqref{eq:w-t1-t2-def}.

Since the  electron-spin operator $\vec{S}$  is the generator 
of the electron-spin rotation, one arrives in the high-temperature limit at
\begin{align}
P_{g_0,g_0}(T,\tau) = \frac{1}{2}-8\braket{S_yS_z(T-\tau)S_yS_z(-T-\tau)}.
\label{eq:Pg0g0-SzSy}
\end{align}
Incorporating $S_y$ as well as $S_z$, $P_{g_0,g_0}(T,\tau)$ 
differs from  $w(t_1, t_2)$ introduced in the previous section.
But with an external magnetic field applied only in the
$x$-direction, the system is invariant under rotation in the $y-z$-plane.  Therefore, we expect both fourth-order correlation functions $P_{g_0,g_0}$ and $C_4$ to exhibit similar properties.

A different type of fourth-order spin correlation function has been
investigated
by Li and Sinitsyn  \cite{LiSinitsyn2016} using a classical approach. 
These authors target the cross correlations between the 
spin-noise power at different frequencies. The 
higher-order spin noise function factorizes into the product
$C_2(\w_1)C_2(\w_2)$  
for uncorrelated frequencies
while the cumulant reveals cross correlations of the different frequency components.

In this paper, however, we focus on the two fourth-order correlation functions 
defined in the time domain as derived above that are directly linked to recent experiments.

\section{Models}
\label{sec:theory}

\subsection{Central-spin model (CSM)}
\label{sec:CSM}

In an InGaAs quantum dot charged
with a single electron, the hyperfine interaction between the electron spin and the nuclear spins dominates the short-time dynamics.   The Hamiltonian $H_{\text{CSM}}$ describes the simple central-spin model
\begin{align}
H_{\text{CSM}}= g_{\text{e}}\mu_{\text{B}}\vec{B}_{\text{ext}}\vec{S}+
\sum_{k=1}^N  \tilde{A}_k \vec{I}_k\vec{S}
\end{align}
where the electron spin $\vec S$ interacts via hyperfine interaction with the surrounding $N$
nuclear spins $\vec{I}_k$ and precesses in the external magnetic field $\vec{B}_{\text{ext}}$. 
The hyperfine coupling constants $\tilde A_k$ are proportional to the probability of the electron at the position of the $k^{\text{th}}$ nucleus $|\psi(\vec{R}_k)|^2$. 
In a QD, the sum
\begin{eqnarray}
\tilde A_s  &=& \sum_{k=1}^N  \tilde{A}_k
\end{eqnarray}
is a universal constant due to the normalization of the wave function
and independent of the shape of the wave function.

Since we only investigate a negatively charged
quantum dot, the hyperfine interaction is isotropic. For hole doped QDs 
the hyperfine interaction acquires an anisotropy defined by the growth direction \cite{Testelin2009}. 
The stable isotopes of Arsenic and Gallium have a nuclear spin of 3/2, and the stable isotopes of Indium have a nuclear spin of 9/2. 
For simplicity we take all nuclear spins $I_k$ to be 3/2.

The fluctuation of the Overhauser field
\begin{align}
\vec{B}_N = \sum_{k=1}^N \tilde{A}_k \vec{I}_k
\end{align} 
is given by
\begin{align}
\omega^2_{\text{fluc}}=\frac{4I(I+1)}{3} \sum_{k=1}^N  \tilde{A}_k^2,
\end{align}
defining the energy scale of the electron spin's decoherence in the quantum dot induced by the hyperfine interaction.
The time scale $T^{\ast}=1/\omega_{\text{fluc}}$
is used as the natural time unit throughout the paper. In experiment, 
typical values of 1-3\,ns are found for $T^{\ast}$ in QDs, depending on their lateral size.

With the dimensionless hyperfine coupling constants
\begin{align}
A_k=\frac{\tilde A_k}{\omega_{\text{fluc}}}
\end{align}
and the dimensionless external magnetic field
\begin{align}
\vec b = \frac{g_{\text{e}}\mu_{\text{B}}}{\omega_{\text{fluc}}}\vec{B}_{\text{ext}},
\end{align}
the Hamiltonian takes the form
\begin{align}
H_{\text{CSM}}=\omega_{\text{fluc}}\left(\vec b +\sum_{k=1}^N  A_k \vec{I}_k\right)\vec S = \omega_{\text{fluc}}\left(\vec b + \vec b_N\right)\vec S
\end{align}
with $\vec b_N$ being the dimensionless Overhauser field. 
While this CSM describes short-time spin dynamics $t \approx T^{\ast}$ very well,
additional interaction terms are required, such as the Zeeman splitting of the nuclei and the quadrupolar interaction, 
in order to make contact to spin noise experiments \cite{PhysRevLett.115.207401,Glasenapp2016}.
The magnetic dipole-dipole interaction between
neighboring nuclei in GaAs can be neglected in our calculations
since its strength is about $(100 \upmu s)^{-1}$ for
two neighboring nuclei and decays as $1/r^3$ as has been pointed out in the 
review \cite{HansonSpinQdotsRMP2007}.

 Different distributions have been used in model calculations \cite{PhysRevB.89.045317, CoishLoss2004, PhysRevB.88.085323} for the CSM. 
The short-time spin decay, however, is universal and independent
of the detailed shape of the distribution function \cite{Merkulov2002}
and only determined by $\omega_{\text{fluc}}$.
Only the long-time asymptotic \cite{CoishLoss2004}
of $C_2(t)$ depends on the distribution function $P(\tilde A_k)$
in the absence of or at very small external magnetic fields. For large electric Zeeman energy, $|\vec{b}| \gg 1$,
the higher momenta of $P(\tilde A_k)$ are known to be not of importance for the spin noise spectrum \cite{Glazov2012,Glasenapp2016}.

Here, we have used the exponentially decaying
coupling constants defined by
\begin{align}
\label{eq:ak-alpha}
\tilde A_k(\alpha) = A_{\text{max}}\text{e}^{-\alpha(k-z_k)/N}
\end{align}
By setting $z_k=0$, and $\alpha=1$,
we recover the exponential distribution of the coupling constants describing a Gaussian electron wave function in a two-dimensional QD as introduced by Coish and Loss \cite{CoishLoss2004}.  In a real material, the $10^5$ active nuclear
spins generated an almost continuous distribution of $\tilde A_k$. In order to mimic such a continuum
we resort to the so-called z-averaging introduced by Yoshida et al \cite{YoshidaWithakerOliveira1990} in the context
of the numerical renormalization group \cite{AndersSchiller2006,BCP08}. By generating configurations with $z_k$ determined from an uniform distribution $z_k \in [-0.5, 0.5]$, and averaging over different configurations, the averaged discrete 
spectrum of a Hamiltonian approaches that of a continuum for large numbers of configurations. 
Configuration averaging has been successfully
employed in numerical 
simulation \cite{PhysRevB.89.045317,PhysRevLett.115.207401,Glasenapp2016}
of the spin noise to minimize the finite size effects in the calculation.
For N=5, we have averaged over $N_C=32$ different
configurations of $\{\tilde A_k\} $. For N=6 and N=7,  $N_C=16$ 
configurations were sufficient to lessen finite size noise. 
The parameter $\alpha$ governs the 
ratio between the largest and the smallest hyperfine coupling. The methods we employ to model the system limit the bath size severely, see Sec.\  \ref{sec:methods}. If $\alpha$ is too large, the weakly coupling nuclear spins don't contribute to the spin dynamics, effectively decreasing the bath size further.  Therefore, we choose $\alpha=0.5$ instead of $\alpha=1$.
We illustrate the marginal difference between setting $\alpha=0.5$ and $\alpha=1.0$
in Fig,\ \ref{c2_Ik-qdiff} found in in Sec.\ \ref{sec:Hq-C2}.

\subsection{Nuclear-electric quadrupolar  interaction}
\label{sec:Hq}

When a QD grows on a substrate, lattice strain can cause the nuclei to take on a prolate charge distribution, which presents an electric quadrupolar moment. This plays a central role in the nuclear spin dynamics of QD \cite{Bulutay2012}. The quadrupolar interaction term \cite{PhysRev.79.685,Slichter1996}
\begin{align}
\begin{split}
\label{eqn:Hq}
H_Q =& \sum_{k=1}^N \tilde q_k\left((\vec{I}_k\vec{n}_k^z)^2 - \frac{I(I+1)}{3}\right.\\
&+\left. \frac{\eta}{3}(\vec{I}_k\vec{n}_k^x)^2-\frac{\eta}{3}(\vec{I}_k\vec{n}_k^y)^2\right)
\end{split}
\end{align} 
originates from the interaction of an stress induced electric field gradient with the quadrupolar moment of the nucleus
and its overall strength at the kth nucleus is denoted by $ \tilde q_k$.
The anisotropic factor $\eta$ been found to be $\eta\approx 0.5$ \cite{Bulutay2012}. Here it is taken
as independent from the nucleus for simplicity.
Note, that in contrary to simplifications \cite{Sinitsyn2012} $H_Q$ conserves time reversal symmetry \cite{PhysRevLett.115.207401}.
The distribution of spatial orientations of the unit vector $\vec{n}^z_k$ defining the local nuclear easy axis
are dependent on the underlying material, and the corresponding unit vectors $\vec{n}_k^x, \vec{n}_k^y$ 
complete the local orthonormal coordinate system at the nucleus $k$.

The distribution of the local easy axis directions
is determined by the QD growth. The microscopic details of
anisotropy factor $\eta$ as well 
as the distribution of the deviation angle $\theta$ of the easy axis
from the z-direction, i.\ e.\ $\cos \theta= \vec{n}_k^z \vec{e}_z$,
have been investigated by Bulutay et al. \cite{Bulutay2012,Bulutay2014}. We have used 
the mean deviation angle of $\theta=23^{\circ}$ reported for a typical GaInAs QD
by generating isotropically distributed vectors $\vec{n}_k^z$
and discarding any vector at an angle from the $z$-axis larger than $\theta_{\text{max}}=34^{\circ}$
\cite{PhysRevLett.115.207401}.

In order to eliminate finite size effects introduced by different
numbers of nuclear spins, the ratio 
\begin{align}
Q_r = \frac{1}{\tilde A_s} \sum_{k=1}^N \tilde q_k.
\end{align}
turned out \cite{HackmannPhD2015,PhysRevLett.115.207401,Glasenapp2016} 
to be a useful measure of the relative  quadrupolar interaction strength.
We explicitly demonstrate this  in fig.\ \ref{c2_ndiff} below.
In a first step, we randomly determine  $c_k$
from a uniform distribution $c_k \in [0.5:1]$
and then
calculate $Q' =\sum_k c_k$ in order to obtain the coefficients 
$\tilde q_k = c_k \tilde A_s Q_r/ Q'$ entering eq.\ \eqref{eqn:Hq}  \cite{Bulutay2012,Bulutay2014,PhysRevLett.109.166605,HackmannPhD2015, PhysRevLett.115.207401,Glasenapp2016}.

\subsection{Nuclear Zeeman effect}

While  for small magnetic fields the nuclear Zeeman term can be neglected \cite{Merkulov2002,PhysRevB.89.045317,Glasenapp2016},
the nuclear Zeeman energy,

\begin{align}
H_{\text{Z}} =  \sum_{k=1}^N g_{\text{k}} \mu_{\text{k}}\vec{I}_k \vec{B}_{\text{ext}} = \omega_{\text{fluc}} \sum_k z_k \vec{I}_k \vec{b},
\end{align} 

has to be included when its value becomes comparable to the hyperfine interaction. 
Its relative magnitude with respect to the electronic Zeeman energy is determined by the ratio

\begin{align}
z_k = \frac{g_\text{k}\mu_\text{k}}{g_{\text{e}}\mu_{\text{B}}}.
\end{align}

Realistic materials consist of different elements as well as different isotopes 
all having individual factors $z_k$, even if the nuclear spin length is $I=3/2$ for all stable
Ga and As isotopes. 
While the main effect of the nuclear-electric  quadrupolar
interaction is to provide an additional dephasing mechanism due
to random orientation of the two time reversal doublets 
formed by the four states of an $I=3/2$ spin, 
an external magnetic field lifts these degeneracies
and suppresses spin-flip processes between the electronic spin and the nuclear spin bath.
The value of $z_k$ defines when this crossover sets in. Note that the variation of $z_k$ is of the order of 30\%. Therefore, different values of $z_k$
do not change this fundamental mechanism but extends the crossover region, and will indeed
have a small influence of the overall long-time decay rates: the larger the average 
$z_k$, the earlier this effect sets in and we expect a lower time scale $T_2$.

This paper, however, only targets the fundamental
understanding of the origin of the different long-time scales seen.
Given the limited bath size we can simulate, different kinds of $z_k$ additionally
lead to an increase of finite size effects. To better model a QD with a large number of nuclei,  we use a uniform $z_k=z$ given by an average over all nuclei. For an InGaAs QD, we set
$z \approx 1/800$.  The experiment by Bechtold et al.\ \cite{Bechtold2016} is performed with external magnetic fields up to $B_x=$4\,T. %
Since $T^{\ast}\approx 1\,\text{ns}$, 4\,T corresponds to  $b_x = 200$ and $z b_x$ becomes comparable to the 
 magnitude of the hyperfine interaction and is therefore non-negligible.

\section{Methods}
\label{sec:methods}

\subsection{Semi-classical approximation (SCA)}

The period of the electron spin precession ($2\pi T^{\ast}$) in the hyperfine field of the nuclei is found to be $\mathcal{O}(\text{ns})$, while the precession period of a nuclear spin in the hyperfine field of the electron is $\mathcal{O}(\upmu\text{s})$. In the short time range, the fast precession electron spin sees a frozen 'snapshot' of the Overhauser field. Since the number of nuclei present in the QD is of the order $\mathcal{O}(10^5)$ \cite{Glasenapp2016, Merkulov2002}, this 'snapshot' follows a Gaussian distribution. In the SCA the Overhauser field is taken as a static variable and
an ensemble average is performed over the central spin precession in an effective magnetic field given by the sum of the external magnetic field and the Overhauser field. The method was used to analyze 
the decay of the central spin
by Merkulov et al.\ \cite{Merkulov2002}. We will see that the SCA accurately describes short time dynamics in the order of $T^{\ast}$, but not the long time dynamics $t \gg T^{\ast}$.

\subsection{Exact diagonalization (ED)}
\label{sec:ED}

Since the SCA treats the Overhauser field statically, and includes neither the quadrupolar coupling nor the nuclear Zeeman term, we also employ ED of the Hamiltonian to analyze the effects of interactions beyond the hyperfine coupling on
the long-time dynamics. \\
Let $E_k$ be the eigenenergies and $\ket{k}$ the eigenvectors of the Hamiltonian
\begin{align}
H\ket{k}=E_k\ket{k}
\end{align}
and any time dependent operator $\hat O(t) = U^{\dagger}(t)\hat O(0) U(t)$ with the time evolution operator $U(t)=\exp(-\text{i} Ht)$. 
A fourth-order autocorrelation function of $\hat O$ can be expressed as
\begin{align}
\begin{split}
\braket{\hat O(t_1)\hat O(t_2)\hat O(t_3)\hat O(0)} =& \frac{1}{D} \sum_{ijkl} O_{ijkl} f_{ijkl}(t_1, t_2, t_3)
\end{split}
\end{align}
where the time dependency is accounted for by the factor
\begin{align}
f_{ijkl}(t_1, t_2, t_3) = \text{e}^{\text{i}(E_i-E_j)t_1}\text{e}^{\text{i}(E_j-E_k)t_2}\text{e}^{\text{i}(E_k-E_l)t_3}
,
\end{align}
and
\begin{align}
O_{ijkl} = \braket{i|O|j}\braket{j|O|k}\braket{k|O|l}\braket{l|O|i}
\end{align}
denotes the product of matrix elements of the operator $\hat O$.
Although it is straight forward to evaluate this expression exactly in a finite size system, the sum over four indices running over the dimension of the Hilbert space turns out to be the limiting factor and restricts us to small bath sizes $N$. Since we have the semi classical approximation at hand, which includes the proper limit $N\to \infty$, we can gauge the quality of our ED by a direct comparison of the results when restricting ourselves to $H=H_{\text{CSM}}$.

\subsection{Lanczos Algorithm with restart}
\label{sec:lanczos-restart}

The number of nuclear spins that can be included in the ED is severely limited.
In order to make progress to access large number of $I=3/2$ nuclear spins,
we used the stochastic evaluation of the trace
\cite{RevModPhys.78.275,PhysRevB.89.045317}
by averaging over a small number $R \ll D$ of randomly chosen states $\ket{r}$ . The relative error made by the stochastic evaluation of the trace
\begin{align}
\text{Tr}(O) \approx \frac{1}{R} \sum_{r=0}^{R-1} \braket{r|O|r}
\end{align}
is of the order $\mathcal{O}(1/\sqrt{DR})$ \cite{RevModPhys.78.275}. 

Using this technique, one can approximate the fourth-order autocorrelation
function  for equidistant laser pulses $G_4(t, 2t, t)$ (eq. \eqref{eqn:g4}) by
\begin{align}
\label{eqn:35}
G_4(t, 2t, t) = \frac{2}{RD}\sum_r^{R}\braket{1_r(t)|P_{\downarrow}e^{\mathrm{i}Ht}P_{\downarrow}e^{-\mathrm{i}Ht}P_{\downarrow}|2_r(t)}
\end{align}
where
\begin{eqnarray}
\ket{1_r(t)} &=& e^{-\mathrm{i}H t} \ket{r} \\
\ket{2_r(t)} &=& e^{-\mathrm{i}H t} P_\downarrow \ket{r}.
\end{eqnarray}

In order to proceed, we need to track the time evolution of quantum states for very long time
up to a few $\upmu$s $\approx 5000\,T^{\ast}$. 
Since this is very challenging for any  polynomial  approach,
we discretize the time $t_n=n\tau, n\in \mathbb{Z}$ and use the recursion relation
\begin{eqnarray}
\label{eq:psi-time-prop}
\ket{\psi(t_n)} &=& e^{-\mathrm{i}H \tau} \ket{\psi(t_{n-1})} 
\end{eqnarray}
to propagate a state $\ket{\psi(t)}$ in reasonably small time steps $\tau$.

There are several options such as a Chebychev polynomial approach
\cite{RevModPhys.78.275,PhysRevB.89.045317}
of a Runge-Kutta algorithm for performing each recursion step.
Here we employed the Lanczos-Krylov Algorithm \cite{Lanczos50aniterative} for calculating
the real-time propagation. The Krylov space spanned by the series of non-orthogonal
vectors
\begin{align}
\ket{\psi}, H\ket{\psi},, H^2\ket{\psi},, ..., H^{M-1}\ket{\psi},
\end{align}
defines a $M$ dimensional subspace of the Hilbert space 
where $\ket{\psi}$ denotes the starting vector of the approach \cite{Schmitteckert2004,SaadSparseLinearSystemsBook2003,Hanebaum2014}.
For evaluating the  time evolution in \eqref{eq:psi-time-prop}, the starting vector 
would be given by $\ket{\psi_0}= \ket{\psi}=\ket{\psi(t_{n-1})}$.

In a first step, the next orthonormal basis vector is constructed via
\begin{eqnarray}
b_1 \ket{\psi_1} &=& H \ket{\psi_0} - a_0 \ket{\psi_0}
\end{eqnarray}
such that $a_0 = \bra{\psi_0}H \ket{\psi_0}$ and $b_1$ is obtained from the normalization of 
$\ket{\psi_1}$. This leads
to the Lanczos recursion relation \cite{Lanczos50aniterative,SaadSparseLinearSystemsBook2003}
\begin{eqnarray}
b_{n+1} \ket{\psi_{n+1}} &=& H \ket{\psi_n} - a_n \ket{\psi_n} -b_n\ket{\psi_{n-1}} 
\end{eqnarray}
providing a 
$M\times M$ dimensional tridiagonal hermitian Hamilton matrix 
$\doubleunderline{H(M)}$,
with the diagonal elements $H_{nn}= a_n$ and the off-diagonal part $H_{nn+1}= b_{n+1}$, where
the recursion is stopped after $M-1$ steps. The Lanczos states $\{ \ket{\psi_n} \}, n=0,\cdots M-1$ 
serve as a  complete orthonormal basis set of $M$-dimensional Krylov space.

For the Lanczos method with restart,
the state $\ket{\psi(t_{n-1})} $ from the previous time step is used
as the new starting vector $\ket{\psi_0}$ in the algorithm above.  
Through diagonalization of $\doubleunderline{H}(M)$, the eigenstates $\ket{\nu^n}$
\begin{align}
\underline{\underline{H}}(M)\ket{\nu^n}= \epsilon_n\ket{\nu^n}
\end{align}
and the corresponding eigenvalues $\epsilon_\nu^n$ are obtained. It allows to
construct an approximate solution of the real-time evolution of  $\ket{\psi(t_{n-1})}$ according
to Eq.\ \eqref{eq:psi-time-prop}:
\begin{eqnarray}
\ket{\psi(t_n)} &=& \sum_{n=0}^{M-1} \mathrm{e}^{-\mathrm{i}\epsilon_\nu^n\tau}
c_n(t_{n-1})  \ket{\nu^n}.
\end{eqnarray}
where $c_n(t_{n-1})  = \braket{\nu^n|\psi(t_{n-1})}$. This expression becomes exact in the limit $M\to \infty$ or
$\tau\to 0$ for a finite $M$. For the calculation below, we typically have used
$M=400$ and a propagation time of $\Delta t =t_n-t_{n-1}= 2/\omega_{\rm fluc}$
for high magnetic fields and  $M=200$ and $\Delta t =100/\omega_{\rm fluc}$ for $\vec b = 0$, 
before we start the next Lanczos
step using $\ket{\psi}=\ket{\psi(t_{n})}$. This provides a numerically expensive but very precise way of
propagating quantum states for very long times.

After this excursion on the details of the Lanczos algorithm, we come
back to the original challenge to track the time evolution in $G_4(t,2t,t)$ stated in
Eq.\ \eqref{eqn:35}. In a first step, we
arrive at the approximate representation of the state
\begin{align}
\ket{1_r(t_n)} = \mathrm{e}^{-\mathrm{i}H\tau}\ket{1_r(t_{n-1})}=\sum_{n=0} ^{M-1}  \mathrm{e}^{-\mathrm{i}\epsilon_\nu^n\tau}
c^1_n(t_{n-1})  \ket{\nu^n}.
\end{align} 
employing the Lanczos algorithm with restart
where $c^1_n(t_{n-1})  = \braket{\nu^n|1_r(t_{n-1})}$. 
To arrive at $\ket{1_r(t_n)}$, $n$ such Lanczos time evolution steps have to be computed.

%
%

The same time evolution is needed for the vector $\ket{2_r(t)}$,
\begin{align}
\ket{2_r(t_n)} = \mathrm{e}^{-\mathrm{i}H\tau}\ket{2_r(t_{n-1})}=\sum_{n=0} ^{M-1} \mathrm{e}^{-\mathrm{i}\epsilon_\nu^n\tau}
c^2_n(t_{n-1})
\ket{\nu^n},
\end{align} 
where $c^2_n(t_{n-1})=\braket{\nu^n|2_r(t_{n-1})}$.

We then repeat the time evolution on the vectors $\ket{3_r(t_n)} = e^{-\mathrm{i}Ht_n} \ket{3_r(0)}$,
$\ket{3_r(0)}=P_{\downarrow}\ket{1_r(t_n)}$,
and 
$\ket{4_r(t_n)}=e^{-\mathrm{i}Ht_n} \ket{4_r(0)}$,  $\ket{4_r(0)}=P_{\downarrow}\ket{2_r(t_n)}$
using the same algorithm and finally express $G_4(t, 2t, t)$ by the matrix element

\begin{align}
G_4(t, 2t, t) = \frac{2}{RD}\sum_r^{R}\braket{3_r(t)|P_{\downarrow}|4_r(t)}.
\end{align}

Since the $P_{\downarrow}$ operator is diagonal in the original Ising basis, 
this expression is easily evaluated. In conclusion, $4n$ Lanczos time evolutions are necessary to arrive at the single value of $G_4(t_n, 2t_n, t_n)$.
We have implemented the calculation of $G_4(t_n, 2t_n, t_n)$ by an massive parallelized 
algorithm for obtaining $\ket{3_r(t_n)}$ and $\ket{4_r(t_n)}$ from an initial set of $N$ vectors $\ket{1_r(t_n)}$
and $\ket{2_r(t_n)}$.

\section{Results}
\label{sec:results}

In this section, we present our results for the fourth-order correlation $G_4$ measured by the three measurement pulse experiment by Bechtold et al.\ \cite{Bechtold2016} and the fourth-order correlation $P_{g_0, g_0}$ measured by Press et al.\ \cite{Press2010}. We analysed the influence of the different interactions present in the QD on the long-time behaviour of  $w(t,t)=G_4(t,2t,t)$. In particular, we can show
that the CSM is insufficient to explain the experimental findings. We need to add the 
nuclear-electric quadrupolar  interaction $H_Q$ as well as the nuclear Zeeman energy to understand the
occurrence of a second, long-time relaxation time $T_2$ as well as its magnetic field dependency.

\subsection{CSM with SCA and ED}
\label{sec:CMS-SCA-ED}

In order to set the stage for the generic behavior of $G_4(t_1,t_1+t_2,t_1)$ in a finite magnetic field
and gauge the quality of a finite size ED calculation, we compare our ED results for the fourth-order correlation functions with those obtained with a SCA using only  $H_{\text{CSM}}$.

Within the SCA, the short-time dynamics of the central spin is modelled
by the precession of a spin in a constant effective magnetic field, 
$\vec{b}_{\rm eff}= \vec{b}+\vec{b}_N = \omega_{\text{L}}\vec{n}$,

\begin{eqnarray}
C_2(t) &=& \frac{1}{4}((n_1^2+n_2^2)^2\cos(\omega_{\text{L}} t)+n_3^2)\\
C_4(t_1, t_1+t_2, t_1)&=& \frac{1}{16}(n_3^4+(n_1^2+n^2_2)^2\cos(\omega_{\text{L}}(t_1-t_2))
\nonumber \\
&& +n_3^2(n_1^2+n_2^2)\left[2\cos(\omega_{\text{L}}t_2)\right.
\label{C4_stuff}
\\
&&+\left. 2\cos(\omega_{\text{L}}t_1)-1-\cos(\omega_{\text{L}}(t_1+t_2))\right].
\nonumber
\end{eqnarray}
and an subsequent averaging over the Gaussian distributed Overhauser fields $\vec{b}_N$  
\cite{Merkulov2002}, where  $\vec n =(n_1,n_2,n_3)$
and  $\omega_{\text{L}}$ 
denotes the corresponding Larmor frequency $\omega_{\text{L}}=|\vec{b}_{\text{ext}}+\vec{b}_N|$. 
Since conservation of energy holds for the individual spin precession in each configuration, 
the picture of pure dephasing without
energy dissipation emerges from the SCA after the configuration averaging.

\begin{figure}[tb]
\begin{center}
\includegraphics[width=0.47\textwidth,clip]{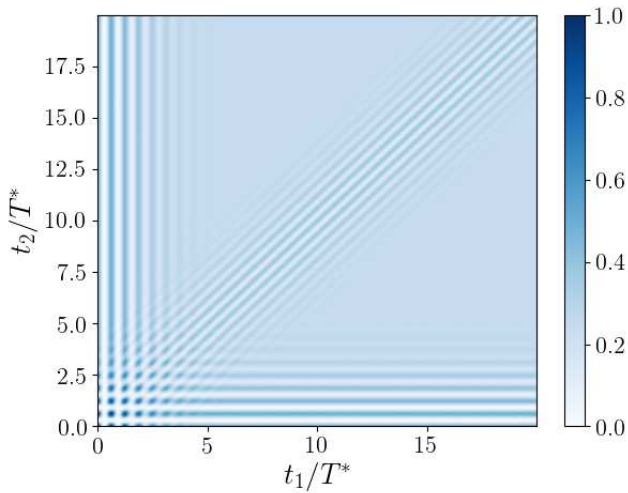}
\caption{SCA of $G_4(t_1, t_1+t_2, t_1)$,  with an external magnetic field of $b_x=10$. It was averaged over $10^6$ Overhauser fields, determined from a Gaussian probability distribution. }
\label{merkfull}
\end{center}
\end{figure}

Fig.\ \ref{merkfull} shows $G_4(t_1, t_1+t_2, t_1)$ averaged over $10^6$ Gaussian distributed Overhauser fields. In this two dimensional color plot, the color encodes the magnitude of the correlation function. In the following, we will focus on two special lines in the $(t_1,t_2)$-plane: The diagonal defined by $t_1=t_2$ and the anti-diagonal for a fixed $T'=t_1+t_2$. For these two cases, there exists published experimental data
\cite{Bechtold2016}.

\begin{figure}[t]
\begin{flushleft}a)\end{flushleft}\vspace{-21pt}
\begin{center}
\includegraphics[width=0.46\textwidth,clip]{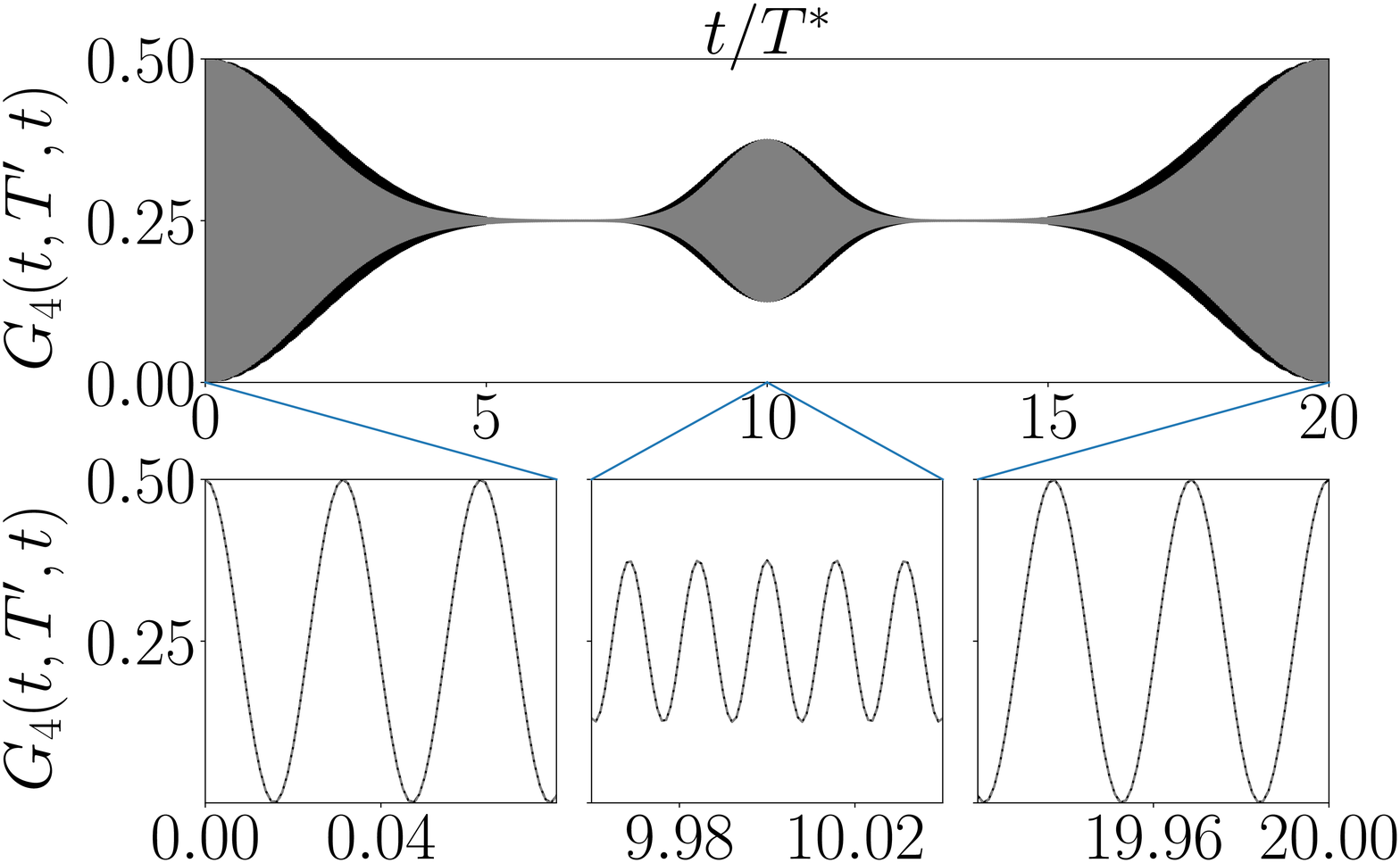}\\
\end{center}
\begin{flushleft}b)\end{flushleft}\vspace{-21pt}
\begin{center}
\includegraphics[width=0.46\textwidth,clip]{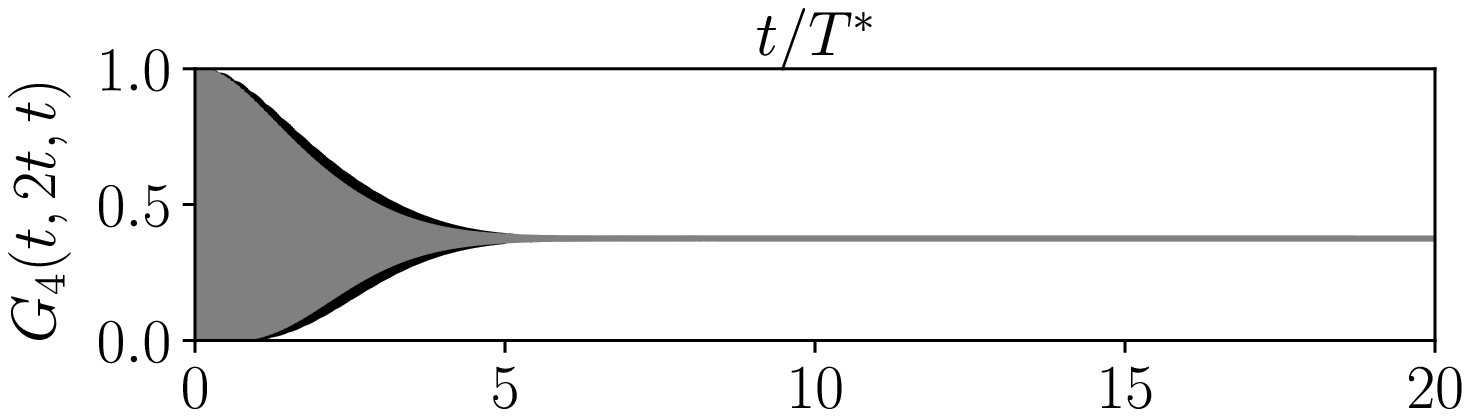}
\caption{Cuts through the SCA (grey) of $G_4(t_1, t_1+t_2, t_1)$ and the exact diagonalization (black) with $N=6$ $\frac{3}{2}$-spins and an external magnetic field $b_x=200$, averaged over $10^6$ Overhauser fields. a) shows the cut where the total time of measurement $t_1+t_2=T'$ is a constant $T'/T^{\ast}=20$. b) shows the cut where the pulses are equidistant, $t_1=t_2$.}
\label{merk_ed_cuts}
\end{center}
\end{figure}

In Fig.\ \ref{merk_ed_cuts}(a), the cut along the anti-diagonal direction, keeping $t_1+t_2=20T^{\ast}$
for $G_4(t,T'=20T^*,t)$ is depicted using both the SCA and the results of an full ED for the  Hamiltonian $H_{\text{CSM}}$. For $t\approx 0$ and $t\approx T'$, oscillations with a frequency of $\omega_{\text{L}}$ 
with a  Gaussian envelope function can be seen, as plotted in the enlarged figures below.

In the middle, $t\approx T'/2$, a frequency doubling with $\omega_{\text{L}}$, also characterized by
a Gaussian envelope function with the same characteristic time scale $T^*$, is observed.
The cause of this frequency doubling can be easily understood. In a finite  external magnetic field, 
$G_4$ is reduced to $G_4(t,T',t) \approx 1/4 +2 C_4(t,T',t)$ according to Eq.\ \eqref{eq:G4-C2-C4}
since $C_2(t)$ completely decays for $t\gg T^*$. 
For very strong external magnetic fields in the $x$-direction, we can additionally
neglect  the Overhauser field in leading order and obtain 
\begin{eqnarray}
\label{eqn:C4-T2}
C_4(t, T, t) &=& \frac{1}{16}\cos(2\omega_{\text{L}}(t-\frac{T'}{2}))
\end{eqnarray}
from Eq.\ \eqref{C4_stuff}. Matching the Gaussian envelope of the ED results in fig. \ref{merk_ed_cuts} to the results of Bechtold et al.\ \cite{Bechtold2016}, we extracted the characteristic timescale as $T^{\ast}=1\,\mathrm{ns}$ which we used in all calculations as reference scale.

In the diagonal cut  depicted in fig.\ \ref{merk_ed_cuts}(b), the correlation function oscillates with the Larmor frequency in the short-time range and  quickly converges to a constant value of $3/8$ on the time 
scale $T^*$. Again, this can be understood by examining Eq.\ \eqref{eq:G4-C2-C4}.
Since $C_2(t)$ decays rapidly to zero, only $C_4(T'/2,T',T'/2)= 1/16$ is remaining according to 
\eqref{eqn:C4-T2} so that $G_4(t, 2t, t)=3/8$ for $t\rightarrow \infty$.

Overall, the agreement of the SCA and the fully quantum mechanical ED with a 
rather small number of $N=6$ nuclear spins is remarkable  in a larger external magnetic field,
$|\vec{b}_{\rm ext}| \gg |\vec{b}_N|$. Finite size effects are small and responsible for the slight
deviation of the ED envelop function compared to the Gaussian of the SCA, which is a static approximation for $N\to \infty$.

Apparently, the CSM Hamiltonian is not adequate for describing the long-time dynamics accurately. 
Bechtold et al.\ \cite{Bechtold2016} have reported that $G_4(t, 2t, t)\to 1/4$ for $t\rightarrow \infty$
at moderate fields. In addition, 
there is a crossover reported to an exponential decay with long-time decay time $T_2\approx 1.4\,\upmu$s.
Therefore additional interaction terms are required to cause the non-linear 
long-time dephasing effects observed in the experiment \cite{Bechtold2016}, in particular for the case $t_1=t_2$, where the experiment reveals additional quantum effects. 

To this end, we propose that by adding 
the nuclear-electric quadrupolar  interaction as well as the nuclear Zeeman term to $H_{\text{CSM}}$
we are able to explain the experimental findings.

\subsection{CSM with quadrupolar interaction}
\label{sec:CMS-HQ}

\begin{figure}[tbp]
\begin{center}

\includegraphics[width=0.49\textwidth,clip]{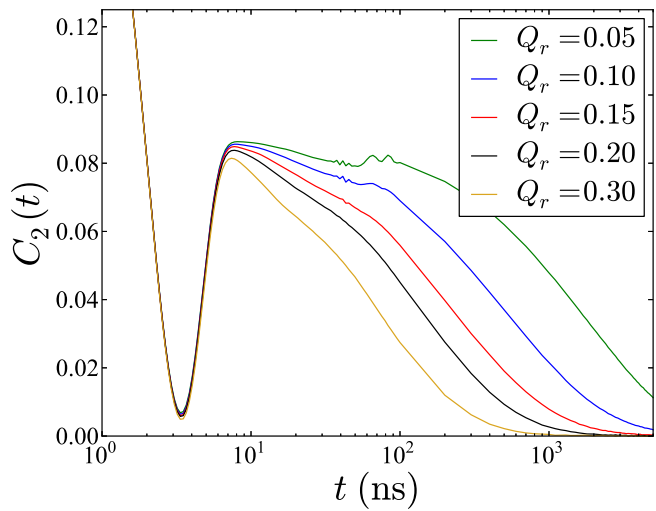}

\caption{$C_2(t)$ computed by Lanczos method with restart with $N=9$ nuclear spins and no magnetic field. The strength of the quadrupolar interaction is varied to determine an experimentally relevant value of $Q_r$.}
\label{c2_qdiff}
\end{center}
\end{figure}

\begin{figure}[tbp]
\begin{center}

\includegraphics[width=0.45\textwidth,clip]{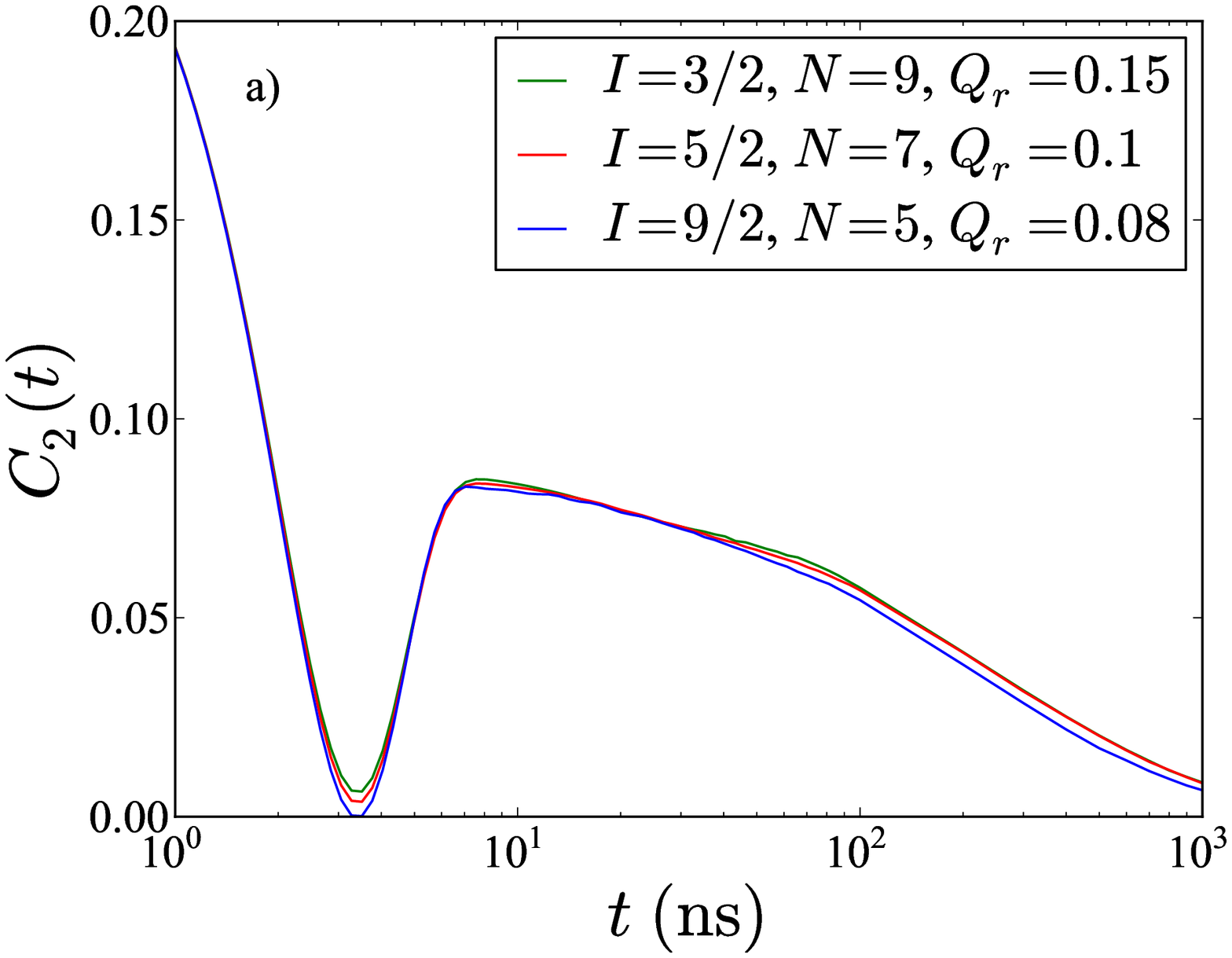}

\includegraphics[width=0.44\textwidth,clip]{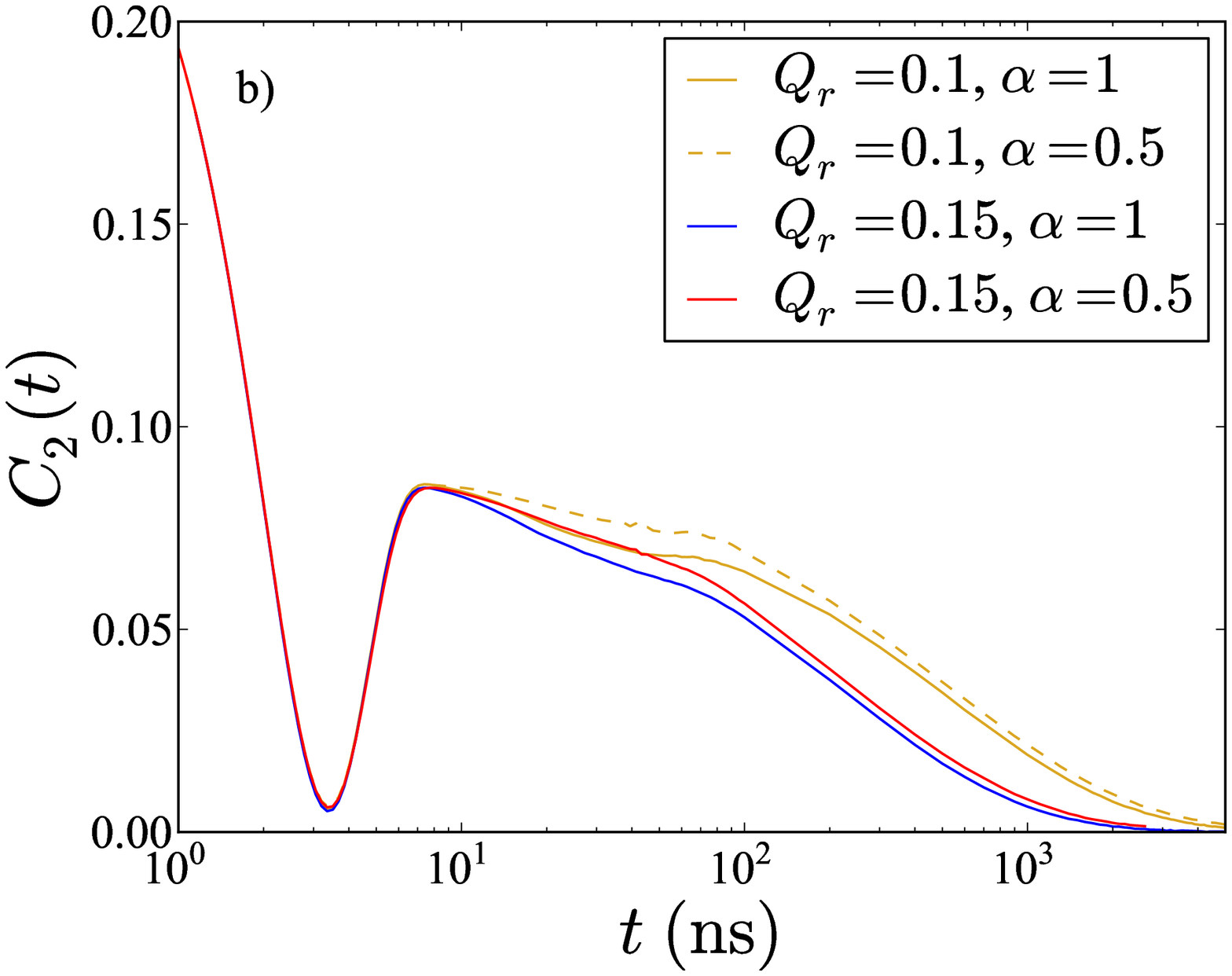}

\caption{(a) $C_2(t)$ computed by Lanczos method with restart for 
different spin length $I=3/2,7/2,9/2$ and an adapted value of $Q_r$.
The numbers of spins have been chosen to keep the Hilbert-space dimension almost identical.
(b)
$C_2(t)$ computed by Lanczos method with restart 
for two different values of $\alpha=0.5,1$ in \eqref{eq:ak-alpha}
and $Q_r=0.1,0.15$ and $N=9$ nuclei with $I=3/2$. Parameters: $\vec{b}=0$.
}
\label{c2_Ik-qdiff}
\end{center}
\end{figure}

Since $H_{\text{Z}}$ suppresses dephasing, 
we neglect this term and only investigate the influence of $H_Q$ at first. 

\subsubsection{The quadrupolar strength $Q_r$}
\label{sec:Hq-C2}

The effect of $H_Q$ is determined by four parameters \cite{Bulutay2012,PhysRevLett.115.207401}: (i) the overall 
quadrupolar strength $Q_r$, (ii) the distribution of the quadrupolar parameter $q_k$, (iii) the anisotropy 
$\eta$ and (iv) the distribution of the local nuclear easy axis all entering $H_Q$ defined in Eq.\ \eqref{eqn:Hq}.
For  (ii)-(iv), we follow Refs.~\cite{Bulutay2012,PhysRevLett.115.207401} by using the parameters 
stated in Sec.\ \ref{sec:Hq} taken for a typical InGaAs QD.

By using the Lanczos approach with restart, we accurately calculated the long-time
behavior of $C_2(t)$ under the influence of the quadrupolar interaction in the absence of an external
magnetic field for a relatively large bath of  $N$=9  nuclear spins with $I=3/2$.
The time evolution is shown  for five different values for  $Q_r$ in fig.\ \ref{c2_qdiff} on a logarithmic
time scale up to $5\,\upmu$s. These Lanczos results reproduce the previous results 
obtained with Chebychev  polynomial approach \cite{PhysRevLett.115.207401}.
We matched  theoretical curves for the spin-spin correlation function  
with the direct measurement of $C_2$ \cite{FinleyNature} and extracted $Q_r\approx 0.15$
for making contact to the experiment. %
This value is very close in magnitude to the parameter used in Ref.\ \cite{Glasenapp2016}
to explain spin-noise data obtained on different  InGaAs QD samples.

\begin{figure}[t]
\begin{center}
\includegraphics[width=0.49\textwidth,clip]{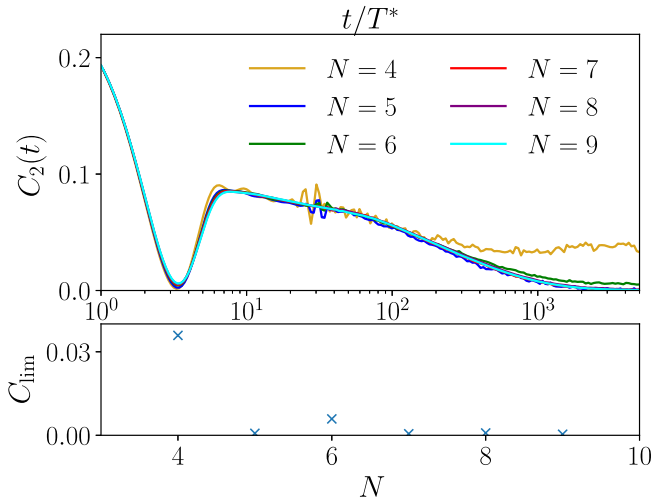}
\caption{$C_2(t)$ with a quadrupolar coupling strength of $Q_r=0.15$, no magnetic field and a varying number of nuclear spins. The long time limit shows an even-odd behavior in regard to bath size, as shown in the lower figure.
}
\label{c2_ndiff}
\end{center}
\end{figure}

In order to determine the finite size effects of our small nuclear spin bath,
we compare $C_2(t)$ for different bath sizes and a fixed quadrupolar coupling of $Q_r=0.15$. 
The top panel of fig.\ \ref{c2_ndiff} demonstrated the fast convergence of $C_2(t)$ with $N$. 
While $C_2(t)$ can be calculated exactly for as large as $N=9$ with a Lanczos method with restart without any problem, this is impossible
for $C_4$ due to the scaling of the nested Lanczos algorithm with the exponential growth of the Hilbert space with $N$.
A finite size analysis for $C_{\rm lim}=C_2(t\to \infty)$ is depicted in the lower panel of fig.\ \ref{c2_ndiff}.
Clearly visible are even-odd oscillations which approach $C_{\rm lim}=0$ at large $N$ within a numerical error of $\mathcal{O}(10^{-4})$. In order to minimize the
finite size effect in the long time limit, we restrict ourselves to odd bath sizes in all following simulations.

Fig.\ \ref{c2_Ik-qdiff}(a) illustrates the  influence of the spin length $I$ on $C_2(t)$. The data for
$I=3/2$ has been taken from Fig.\ \ref {c2_qdiff}. When adjusting the
relative coupling strength $Q_r$ for each $I$ appropriately, we can obtain a universal curve for $C_2(t)$. The small
differences in the minimum are well understood \cite{StanekRaasUhrig2013,HackmannPhD2015} 
 and
are related to the number of spins. This illustrates that the additional dephasing is driven by presence of $H_Q$
while the differences in the spin length are insignificant.

The marginal influence of the choice of $\alpha$ in eq.\ \eqref{eq:ak-alpha}
is depicted in fig.\ \ref{c2_Ik-qdiff}(b). At intermediate times, small deviations are observable, but the long
time limit is not effected and fully determined by the value of $Q_r$.

\subsubsection{Influence of $H_Q$ on $G_4$}

\begin{figure}[bt]
\begin{center}
\includegraphics[width=0.49\textwidth,clip]{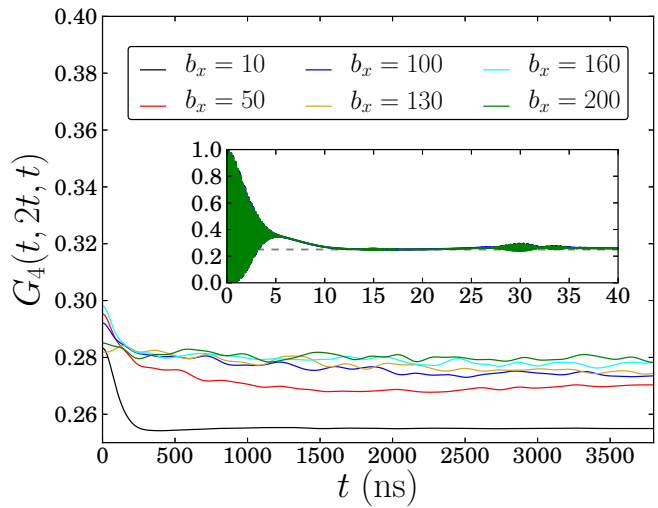}
\caption{$G_4(t, 2t, t)$ 
as function of $t$ for different external magnetic fields.
The curves were smoothed through a low-pass filter to suppress finite size oscillations. 
The inset reveals the short-time behavior
without the low-pass filter. One can see that $G_4$ decays down to a constant level close to 1/4 in $\mathcal{O}(\text{ns})$, and that the time frame of the decay is not dependent on the applied magnetic 
field.
Parameters: $N=5$ with $I=\frac{3}{2}$ and $Q_r=0.15$.
}
\label{csm+q}
\end{center}
\end{figure}

Fig.\ \ref{csm+q} shows $G_4(t, 2t, t)$ with a quadrupolar interaction strength $Q_r=0.15$. 
With this quadrupolar interaction strength, $G_4(t, 2t, t)$ 
rather rapidly approaches a constant, augmented with some finite-size oscillations.
As depicted in the inset of fig.\ \ref{csm+q},  this decay occurs one a time scale of 
approximately 10\,ns. The decay is not 
influenced  significantly by the externally applied magnetic field as long as the magnetic field exceeds $b_x=50$. 
The long-time exponential decay reported in the experiment \cite{Bechtold2016}
is absent in our calculations.

\begin{figure}[t]
\begin{center}
\includegraphics[width=0.49\textwidth,clip]{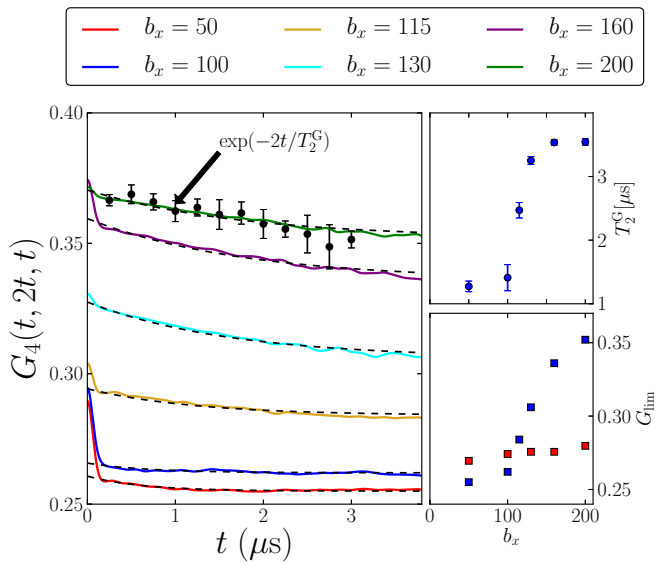}
\caption{$G_4(t, 2t, t)$ 
as function of $t$ for different external magnetic fields.
and a relative strength of the Zeeman splitting of $z=1/800$.  The decoherence time $T^{\text{G}}_2$ and the convergence level $G_{\text{lim}}$ are plotted against the external magnetic field $b_x$. A low pass filter was utilized to suppress finite size oscillations. $G_4(t, 2t, t)$ obtained by a Lanczos algorithm and a bath size of $N=7$ and $b_x=200$ is shown in black dots. 
In the upper right panel , the decay time $T^{\text{G}}_2$ is plotted in blue dots. The long-time limit $G_{\text{lim}}$ is shown in the lower right. The blue and red squares denote $G_{\text{lim}}$ for $z=1/800$ and $G_{\text{lim}}$ for $z=0$, respectively. The dashed lines show the ED data fitted with eqn.\ \eqref{eqn:g4exp}.
Parameters: $N=5$ with $I=\frac{3}{2}$ and $Q_r=0.15$.}
\label{csm+q+ks}
\end{center}
\end{figure}

To suppress finite-size oscillations in favor of showcasing the long-time behavior, we smooth the curves through convolution with a Gaussian function
\begin{align}
g(t) = \frac{1}{\sqrt{2\pi\sigma^2}}\exp{\left(-\frac{t^2}{2\sigma^2}\right)}.
\end{align}
In both fig.\ \ref{csm+q} as well as fig.\ \ref{csm+q+ks}, a standard deviation $\sigma=60\,\text{ns}$ was used for this low-pass filter. This is sufficiently small to not distort the curve progression in $\mathcal{O}(\upmu \mathrm{s})$.

\subsection{Combining $H_{\rm CSM}$ with quadrupolar interaction and Zeeman splitting of the nuclei}
\label{Finley_full}

In order to make a connection to the experiments \cite{Bechtold2016}, a
suppression of the additional long-time dephasing mechanism introduced by $H_Q$ is needed.  
This effect is provided by the nuclear Zeeman term, which dominates the local nuclear spin 
dynamics at large external magnetic fields: The Zeeman splitting of the nuclei suppresses nuclear spin flip processes once the nuclear Zeeman energy is within the order of magnitude of the 
hyperfine interaction.


To this end, we consider the full Hamiltonian $H=H_{\text{CSM}}+H_Q +H_\text{Z}$ in this section. 
With this additional term, a completely different behavior 
for the long-time limit of $w(t,t)= G_4(t, 2t, t)$
emerges as depicted in fig.\ \ref{csm+q+ks}.
In the larger left panel of the figure,  $G_4(t, 2t, t)$ is plotted up to 4$\,\upmu$s for various
magnetic field strengths, again using the low-pass filter to suppress the finite size noise in the long time
evolution. Setting $g=0.55$ for the electron spin in a QD, and the time scale $T^*=1\,$ns,
the dimensionless fields $b_x=50,100,115,130,160,200$ translate to physical units of 
$B_x=|\vec{B}_{\rm ext}|= 1.03, 2.07, 2.38, 2.68, 3.31, 4.14\,$T.  
All curves are calculated using  ED with only $N=5$ nuclear $3/2$-spins, a uniform
average ratio $z=1/800$ and averaged over $N_C=32$ configurations of $\{ A_k\}$.

For low magnetic fields $b_x=50,100$, the results are very similar to the results without $H_Z$ depicted in fig.\ \ref{csm+q}. Rather rapidly, $G_4(t, 2t, t)$ decays to its asymptotic
magnetic field dependent long time limit defined as
\begin{align}
G_{\text{lim}}(b_x) = \lim_{t\rightarrow \infty} G_4(t, 2t, t).
\label{eqn:G4-long-time}
\end{align} 
Here, however $G_{\text{lim}}(b_x=50)$ is lower than the value without nuclear Zeeman term, indicating that for small and intermediate fields the non-linear nuclear dynamics governed by
the combination of the nuclear Zeeman term and nuclear quadrupolar interaction
yields an overall long time decay closer to the theoretical lower limit of $1/4$ for a complete dephasing
of the fourth-order spin correlation functions $C_4(t,2t,t)$.

$G_{\text{lim}}(b_x)$ is shown as red squares for $H=H_{\text{CSM}}+H_Q$
and as blue squares for the full Hamiltonian  in the lower right panel of fig.\ \ref{csm+q+ks}.
A qualitatively different picture emerges once $b_x$ exceeds $100$. While $G_{\text{lim}}$ remains almost magnetic field
independent for $H_{\text{CSM}}+H_Q$, 
$G_{\text{lim}}(b_{\rm ext})$ monotonically increases with $b_x$ for the full Hamiltonian,
as plotted in the lower right panel of  fig.\ \ref{csm+q+ks}. The increase occurs rather rapidly and approaches a
plateau for large fields, since 
$G_{\text{lim}}(b_x)$ cannot exceed the asymptotic value $3/8$ bound by the SCA.

We have analyzed the slow long-time decay of $G_4(t, 2t, t)$ by   
assuming an exponential
form
\begin{align}
G_4(t, 2t, t) = G_{\text{lim}}(b_x) + \xi\exp(-2t/T^{\mathrm{G}}_2)
\label{eqn:g4exp}
\end{align} 
parametrized by the amplitude $\xi$ and the additional decoherence time $T^{\mathrm{G}}_2$
in order to connect to the experimental findings
by Bechtold et al.\ \cite{Bechtold2016}.
These fits are added as dashed lines to the calculated $G_4(t, 2t, t)$ in the left panel of fig.\ \ref{csm+q+ks}.
The decoherence time $T^{\mathrm{G}}_2$ obtained by this fit 
is plotted as function of the external magnetic field in the upper right panel of fig.\ \ref{csm+q+ks}.
$T^{\mathrm{G}}_2$ is small and difficult to determine for small fields and rapidly increases around $b_x=115$ being equivalent to $B_x=2.38\,$T. 
This is in full accordance with Ref.\  \cite{Bechtold2016} where only the data for largest field value of $B=4\,$T 
was fitted with an exponential form. While we find values of $T^{\mathrm{G}}_2\approx 3.5\upmu$s at $B_x=4.14\,$T
by Bechtold et al.\ reported $T^{\mathrm{G}}_2=1.4\pm0.1\,\upmu\text{s}$  
at an external magnetic field of $B_x=4\,$T. 
Note that the experimental data points 
presented in Fig.\ 4(a) of Ref.\  \cite{Bechtold2016} can also be fitted with a larger
$T^{\mathrm{G}}_2$, if one only considers the data points for the long time decay $t>300\,$ns. $T^{\mathrm{G}}_2$ between $2-5\,\upmu$s can be obtained,
indicating that the values for $T^{\mathrm{G}}_2$ strongly depended to the fit procedure.

In order to estimate the finite-size effects, we have calculated the long-time dynamics of $G_4(t, 2t, t)$ 
for $N=7$ nuclear spins using the numerically expensive Lanczos approach with restart  
outlined in 
Sec.\ \ref{sec:lanczos-restart}: The larger the external magnetic field, the larger the number of spins, the larger the
spectrum of $H$, the shorter the time evolution step will be for a given Krylov  space dimension $M$. 
We typically used $M=400$
and a propagation time of $\Delta t = 2\,$ns  for a single step. Since the calculation required two-week runs on our HPC cluster, we have evaluated $G_4(t, 2t, t)$ only at  a set of discrete data points for the largest magnetic field $b=200$.

Within the finite size errors, the Lanczos data  for $N=7$ 
are identical to the results obtained from the ED for $N=5$, leading to the conclusion
that the long-time scale extracted from the ED does not contain substantial finite size errors
for a small increase of $N$.
However, we are aware that the limitation of the energy spectrum of the Hamiltonian introduces finite-size
errors which will influence $G_{\text{lim}}(b_x)$. In the real system, the nearly continuous distribution
function $P(A)$ of the hyperfine coupling will lead to a nearly continuous spectrum of the Hamiltonian
so that phase space for spin-flip processes with be larger, and we expect that long-time
limit $G_{\text{lim}}(b_x)$ will be smaller than in our case. 
We have demonstrated that effect for $C_2(t)$ calculated with the full model
in fig.\ \ref{c2_ndiff}: The finite size offset of $C_2(t\to\infty)$
depicted in the lower panel of fig.\ \ref{c2_ndiff} 
suggests that  a complete decay of $C_2$  can only be achieved 
in the limit  $N\to\infty$.

Since finite-size corrections to asymptotic limit $G_{\text{lim}}(b_x)$
will only influence prefactor $\xi$, the exponential decay time $T_2$ should be  unaltered.
We conclude that our findings for $T^{\mathrm{G}}_2$ agree not only qualitatively but also quantitatively
with the experiments.

\subsection{Spin echo experiments modeled with the full Hamiltonian}
\label{sec:spin-echo}

%

We derived in section \ref{spin-echo} that the amplitude of the spin echo measured by Press et al.\ \cite{Press2010} can be described as a fourth-order correlation function $P_{g_0, g_0}(T,\tau)$. To properly compare our results with the experimental measurements, we study both $P_{g_0, g_0}(T=\text{const},\tau)$ as a function of $\tau$ and the oscillation amplitude of $P_{g_0, g_0}(T,\tau=0)=\bar P_{g_0, g_0}(T)$ as a function of $T$ for several external magnetic fields. Similarly to $G_4(t, 2t, t)$ measured in the Bechtold et al.\ experiment \cite{Bechtold2016}, the amplitude of $P_{g_0, g_0}(T,\tau)$ exhibits a long time exponential decay
\begin{align}
\bar P_{g_0, g_0}(T)=
P_{\text{lim}}+\zeta\exp{\left(-\frac{2T}{T^\text{P}_2}\right)}.
\end{align}

\begin{figure}[tbp]
\begin{center}
\includegraphics[width=0.5\textwidth,clip]{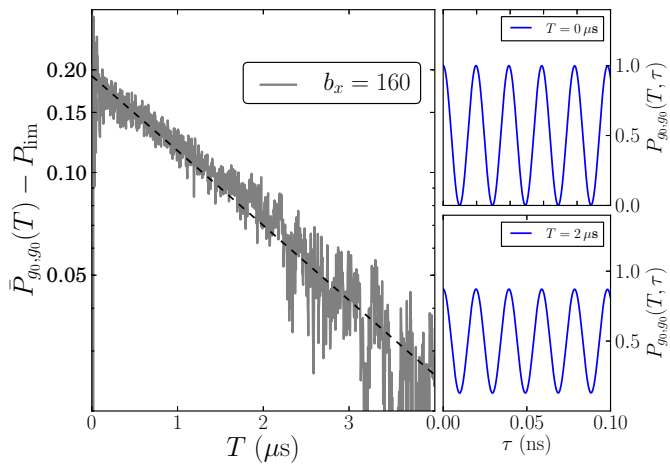}
\caption{Left panel:
Exponential decay of the amplitude of $P_{g_0, g_0}(T,\tau=0)$ with $N=5$, 
here plotted for $b_x=160$ on a log scale, which corresponds  \cite{Press2010} 
to a magnetic field $B_x=4.11\,$T
for $g_\text{e}=0.442$.
The decay time was fitted as $T_2^{\mathrm{P}}=3.96\,\upmu$s. On the right hand side, $P_{g_0, g_0}(T,\tau)$ for a constant pulse distance $T$ is shown, once for $T=0\,\upmu$s and once for $T=2\,\upmu$s. The spin echo oscillates with twice the Larmor frequency, like $G_4(t, T, t)$ at $t\approx T/2$. All system parameters are equal to the ones used in fig. \ref{csm+q+ks} to describe the experiment by Bechtold et al.\
 \cite{Bechtold2016}.
 }
 \label{press1}
\end{center}
\end{figure}

For the computation of $P_{g_0, g_0}(T,\tau)$ using ED, the same system parameters were employed
as for the computation of $G_4(t, 2t, t)$ in fig.\ \ref{csm+q+ks}.
If $T$ is kept constant, $P_{g_0, g_0}(T,\tau)$ oscillates with twice the Larmor frequency
as function of $\tau$. This can be understood analytically:
When neglecting the Overhauser field, one again obtains
\begin{align}
P_{g_0, g_0}(T,\tau) = \frac{1}{2}+\frac{1}{2}\cos(2\omega_{\text{L}}\tau).
\end{align}
The decay of this coherent oscillations of the electron spin in the external magnetic field 
observed in the experiment  \cite{Press2010} is also caused by
the interaction of the electron spin with the surrounding nuclear spins.

\begin{figure}[t]
\begin{center}
\includegraphics[width=0.4\textwidth,clip]{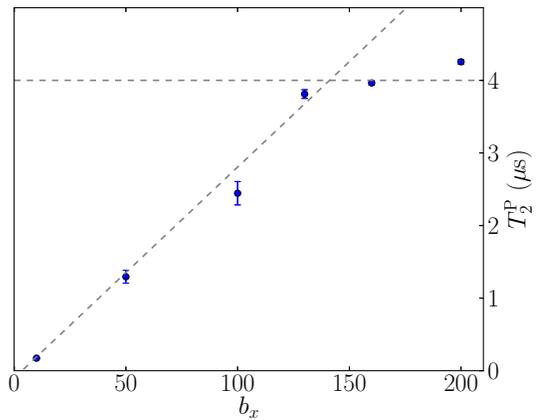}
\caption{The decay time $T^\text{P}_2$ of $\bar P_{g_0, g_0}(T)$ dependent on the external magnetic field $B_x$. The grey dashed lines are guides to the eye, indicating a rising slope that saturates at magnetic fields $b_x>130$. With $g_\text{e}=0.442$ the saturation point is reached at $B_x\geq 3.5\,$T with a saturation level of $T^{\mathrm{P}}_2\approx 4\,\upmu$s. A higher saturation point $B_x=4\,$T at a lower saturation level $T^{\mathrm{P}}_2\approx 3\,\upmu$s was reported by Press et al.\ \cite{Press2010}. }
\label{press2}
\end{center}
\end{figure}

When computing $P_{g_0, g_0}(T,\tau)$ with ED using the full Hamiltonian,
the decay of the spin-echo amplitude is evident, though not as pronounced as in the experiment. Exponential fitting shows that the decay time $T_2$ is similar, but that in our theoretical model the long-time limit of the probability, $P_{\text{lim}}=
\lim_{T\to\infty} P_{g_0, g_0}(T,0)$, lies above the experimental values.
The coherent oscillations in $\tau$ are depicted for two different fixed times $T$
 in the right panel of fig.\ \ref{press1}.
By applying the pulse sequence of $\pi/2-\pi-\pi/2$-pulses long time coherence 
can be maintained on a time scale of $T^\text{P}_2\gg T^*$. We have extracted
$T_2^\text{P}$ as function of the applied external magnetic field shown in
 fig.\ \ref{press2}.
The calculated functional dependence of $T^\text{P}_2\gg T^*$ on the
external magnetic field agrees remarkably well with the experimental data
of fig.\ 4 in ref.\ \cite{Press2010}.

Although  $P_{g_0, g_0}(T,\tau)$ behaves very
similar $G_4(t_1, t_1 + t_2, t_1)$ when identifying $t_1=T+\tau$ and $t_2=T-\tau$,
the magnetic field dependency of the two long-time scales $T^\text{G}_2$ and $T^\text{P}_2$
differs as well as their saturation values, $T^\text{P}_2 \approx 4\,\upmu$s and $T^\text{G}_2 \approx 3.5\,\upmu$s respectively.
This agrees excellently with the experiments,
where the $T^\text{P}_2$ measured by spin-echo method \cite{Press2010}
was also higher than the $T^{\mathrm{G}}_2$ 
obtained by the three measurement pulse experiment \cite{Bechtold2016}. 
We attribute the different in the long-time scales by the difference
of the two different fourth-order spin correlation functions measured in the different
experimental setups: While Bechtold et al.\ have focused on measuring only $S_z$, the $\pi/2$
pulse connect both spin components $S_z$ and $S_y$ orthogonal to the applied
external magnetic field, see eqs. \eqref{eq:G4-C2-C4} and \eqref{eq:Pg0g0-SzSy}.
Note that the underlying dynamics and all model parameters 
have been identical in our calculations for both correlation functions.

\section{Summary and conclusion}
\label{sec:conclusion}

We have shown that the joint probability $w(t_1,t_2)$
of measuring a central spin $\ket{\downarrow}$ twice at times $t_1$ and $t_1+t_2$ in a quantum dot after preparing the central spin in a spin $\ket{\downarrow}$ state is related to the fourth-order correlation function $w(t_1,t_2)=G_4(t_1, t_1+t_2, t_1)$. 
$G_4$ was computed via the SCA, the ED and and the Lanczos algorithm.
To determine the influence of the different interactions, we have studied the long time behavior of $G_4$ with and without nuclear Zeeman splitting and quadrupolar interaction, and have compared the results to the experimental findings reported by Bechtold et al. \cite{Bechtold2016, FinleyNature}. 
$w(t,t)$  exhibits a long time exponential decay in a strong transversal magnetic field 
for equidistant probe pulses $(t_1=t_2)$, which cannot be understood with the semi-classical
approximation or the full quantum dynamics of the CSM including electron and nuclear
Zeeman term.

By including the nuclear-electric quadrupolar interaction in the full-time reversal
form \cite{Slichter1996,Bulutay2012,PhysRevLett.115.207401,Glasenapp2016}
as well as nuclear Zeeman term, for the CSM, the ED produces results which concur qualitatively with the findings of the recent experiments \cite{Bechtold2016, FinleyNature}.  We have used the spin-correlation function $C_2(t)$
to determine $T^*$ and the relative quadrupolar coupling strength $Q_r$ relevant for connection to the experiment.  
 $G_4$ computed using ED for the simple CSM model including only electron Zeeman splitting is very similar to $G_4$ computed by the SCA:
Both approaches perfectly agree in short-time behavior but lack to explain the
experimentally observed long-time decay of the correlation function.
Adding the  nuclear quadrupolar coupling to the Hamiltonian 
and neglecting the nuclear Zeeman term introduces a rapid decay of  $G_4(t,2t,t)$ after the
initial coherent Larmor oscillations on a time scale of  $t_{\rm decay}\approx 10\,$ns.
We have demonstrated that the nuclear Zeeman interaction is essential to observe the crossover
from a fast non-exponential decay caused by the  nuclear quadrupolar coupling in weak 
and intermediate magnetic field to a slow exponential decay at larger magnetic field.
Above a threshold of $b_x\approx 100$, $G_4(t, 2t, t)$ exhibits a long-time exponential decay dependent on the field strength, with a plateauing characteristic time scale $T_2=3.5\,\upmu$s.
With the full model, the increasing external transversal magnetic field $b_x$ leads to a monotonically
increasing asymptotic limit $G_{\text{lim}}(b_x)$. Our findings agree with the experimental results reported by Bechtold et al.\ \cite{Bechtold2016}. 

The intrinsic dephasing time $T_2$ was also measured through the spin echo method by Press et al.\ \cite{Press2010}. We were able to show that the spin echo measurement is described by a fourth-order spin correlation function $P_{g_0,g_0}(T, \tau)$ that is different from $G_4$, involving 
$S_y$ and $S_z$. The decay mechanisms are the same, and the long time behavior 
is qualitatively similar. 
Both $P_{g_0,g_0}$ and $G_4$ reach an asymptotic value for the decay time $T_2$ at large external magnetic fields. The asymptotic decay time of $P_{g_0,g_0}$ is higher, $T^P_2= 4\,\upmu$s, than that of $G_4$, where we have found $T_2^{\mathrm{G}}=3.5\,\upmu$s.
This concurs with the experiments, since Press et al.\ also
reported a higher asymptotic decay time $T_2^{\mathrm{P}}=3\,\upmu$s \cite{Press2010} than Bechtold el al.\ \cite{Bechtold2016}, who measured $T_2^{\mathrm{G}}=1.4\,\upmu$s for $B_x=4\,$T. 
However, 
the approach of the asymptotic value for  large magnetic fields significantly
differs for the two fourth-order correlation functions. While the $T_2^{\text{P}}$ shows a slow linear rise 
-- see also fig.\ \ref{press2} --
that agrees very well with the experiment, $T_2^{\text{G}}$ exhibits a threshold behavior with a rapid rise at $b_x\approx 100$.

For the anti-diagonal of $w(t_1,t_2)$ defined by the fixed distance of
the second probe pulse with respect to the initial pump pulse  ($t_1+t_2=T'$), i.\ e.\ $G_4(t_1,T',t_1)$
both SCA and ED show $G_4$ oscillating with twice the 
Larmor frequency and a Gaussian envelope around $t_1=T'/2$. The cause of the frequency doubling can be analytically understood by  neglecting the Overhauser field compared to
a strong transversal magnetic field in zero order while the Gaussian enveloped is caused
by the dephasing introduced by the Overhauser field.
The same behavior is also observed in $P_{g_0,g_0}(T, \tau)$ at constant $T$.

In conclusion, the CSM model  
augmented with the nuclear-electric quadrupolar coupling and 
nuclear Zeeman splitting is an 
adequate description for fourth-order correlation functions in QD.  
While quadrupolar coupling induces an additional decay of the correlation functions 
towards universal constants, 
a increasing nuclear Zeeman splitting suppresses this effect
leading to a long-time exponential decay. 
Applying the full model predicts a  long-time exponential decay of the fourth-order correlation function  at high transversal magnetic fields 
that  qualitatively and quantitative agrees with the recent experimental results.

\begin{acknowledgments}
For useful discussion we thank M.\
Bayer, A.\ Greilich, J.\ Schnack and T.\ Simmet.
We acknowledge the financial support by the
Deutsche Forschungsgemeinschaft and the Russian Foundation of Basic
Research through the transregio TRR 160.  
\end{acknowledgments}



%

\end{document}